
\magnification=\magstep1
 \baselineskip=0.820truecm
\centerline {\bf LINEAR POTENTIALS AND GALACTIC ROTATION CURVES }
\vskip 1.90truecm
\centerline {\bf Philip D. Mannheim}
\centerline {Department of Physics}
\centerline {University of Connecticut}
\centerline {Storrs, CT 06269-3046}
\smallskip
\centerline{mannheim@uconnvm.bitnet}
\vskip 3.00truecm
\centerline {\bf Abstract}
\vskip 0.60truecm
We derive a simple, closed form expression for the potential of a thin
exponential disk of stars interacting through gravitational potentials of
the form $V(r)=-\beta /r+\gamma r/2$, the potential associated with
fundamental sources in the fourth order conformal
invariant theory of gravity which has recently been advanced by Mannheim
and Kazanas as a candidate alternative to the standard second order
Einstein theory. Using the model, we obtain a reasonable fit (flat to within
$\pm3\%$) to the (prototypical) NGC3198 $HI$ rotation curve data
without the need for any non-luminous or dark matter. Our study suggests that
th
observed flatness of rotation curves may only be an intermediate phenomenon,
and
an asymptotic one.
\vskip 3.00truecm
$~~~~~~~~~~~~~~$October, 1992$~~~~~~~~~~~~~~~~~~~~~~~~~~~~~~$UCONN-92-3
\vfill\eject
\hoffset=0.0truein
\hsize=6.5 truein
\noindent
{\bf (I) Introduction}

In formulating a physical theory it is necessary to both work up from
phenomenological observations and down from fundamental principles, and to
be prepared to revise the insights obtained from both approaches as new
data come on line. However, after a program such as this has been
successfully carried through once, there is then some reluctance on the part
of the
community to have to reopen the issue even in the light of subsequent data.
Consequently, the prevailing view on galactic rotation curve data is that
their deviation from the behavior expected on the basis of the standard
Newton-Einstein theory as applied to the observed galactic luminous matter
surface brightness distribution must
be attributed to a (rather substantial) non-luminous or dark matter
galactic component. Since there is no clear evidence today that the
dominant component of the Universe is in fact non-luminous, there is thus
some merit in going back over familiar ground to see where, if anywhere,
something could be modified.

Noting that there is currently no known theoretical reason which would select
ou
the standard second order Einstein theory from amongst the infinite class of
(al
covariant, metric based theories of gravity that one could in principle
at least consider, Mannheim and Kazanas have reopened the
question of what the correct theory of gravity might be (Mannheim and
Kazanas (1989), Mannheim (1990), Kazanas and Mannheim (1991),
Mannheim and Kazanas (1991),
Mannheim (1992), Mannheim and Kazanas (1992)), and developed an approach which
works down from an additional fundamental principle,
namely that of local scale or conformal invariance, the invariance now believed
possessed by the other three fundamental strong, electromagnetic and weak
intera
This invariance forces gravity to be described uniquely by the fourth order
acti
$I_W = -\alpha \int d^4x (-g)^{1/2}
C_{\lambda\mu\nu\kappa}C^{\lambda\mu \nu\kappa}$
where $C_{\lambda\mu\nu\kappa}$ is the conformal Weyl tensor and $\alpha$
is a purely dimensionless coefficient.
In their original paper Mannheim and Kazanas (1989) obtained the
exact, non-perturbative exterior vacuum solution associated with a static,
spherically symmetric gravitational source in this theory, viz.
$$-g_{00}= 1/g_{rr}=1-\beta(2-3 \beta \gamma )/r - 3 \beta \gamma
+ \gamma r - kr^2 \eqno(1)$$
where $\beta, \gamma,$ and $k$ are three appropriate dimensionful
integration constants. As noted by Mannheim and Kazanas, this solution contains
familiar exterior Schwarzschild solution and thereby yields the standard
exterio
Newtonian potential term and its standard general relativistic corrections
whenever the two additional potential terms in Eq. (1) may be ignored.
The theory thus contains the same solution as the standard Einstein theory
in the appropriate kinematic regime even while not containing the Einstein
equations themselves, this being all that observation can require.
The quadratic term in Eq. (1) may be associated with a general cosmological
background de Sitter geometry and is otherwise uneventful, and thus
(with both the  $3\beta\gamma$ terms being numerically
negligible - see below) the conformal theory
leads to the non-relativistic gravitational potential
$V(r)=-\beta/r+ \gamma r/2 $,
which may then be fitted to data whenever the weak gravity limit is
applicable. $V(r)$ is thus the potential obtained in coming down
from a fundamental principles approach. In this paper we shall study the
implications of this potential by working up from data. As we shall see,
the two procedures even have a chance to converge; however, those
readers who may not be too comfortable with (or even disapprove of) the
whole general conformal gravity program can view this paper purely as an
attempt to identify what phenomenological potentials the currently
available observational data actually permit.

\noindent
{\bf (2) The Potential of an Extended Disk}

If we momentarily ignore all general relativistic questions and first instead
ask simply in what way do data restrict the form of the non-relativistic
gravitational potential, we note immediately that the
validity of the Newtonian $1/r$ potential is so far only definitively
established on solar system or smaller distance scales. There is no such
validation on much larger distance scales as the need for galactic dark
matter clearly attests. Indeed, if galactic data were the only data
available to us, we would not be able to extract out a Newtonian Law at
all. Rather, an implication of galactic rotation curve data is that
something is not falling off at large distances, either the matter
distribution (viz. dark matter) or the law of force itself. (The
possibility that something may in fact be wrong with the standard
gravitational picture has already
been entertained in the literature by Milgrom (1983) and by Sanders (1990)
though from a viewpoint different to the one followed here). As regards
this quite heretical attitude towards the law of force, we would
essentially need to augment the Newtonian term with a second term so as to
keep the good Newtonian results in the appropriate regime; and then the
key issue is finding the class of allowed forms for such an additional term
which are compatible with data. As we shall see, despite the observed
flatness of the curves, a linear two-body potential may also suffice because of
the extended nature of a galaxy.

In order to handle the potential of an extended object such as a disk, many
ways are possible with the most well-known being due to Toomre (1963).
Since the method he developed for the Newtonian case does not immediately
appear to generalize to linear potentials, we have instead developed
an alternate way of handling the $1/r$ case, a method which does readily
generalize. To determine the Newtonian potential of an axially symmetric
(but not yet necessarily thin) surface brightness distribution
$\Sigma(R,z^{\prime})$ of matter sources we need to evaluate the quantity
$$V_{\beta}(r,z)=-\beta\int dR d\phi^{\prime} dz^{\prime} {R\Sigma(R,
z^{\prime}) \over
(r^2+R^2-2rRcos\phi^{\prime}+(z-z^{\prime})^2)^{1/2}} \eqno(2)$$
where $R,~\phi^{\prime},~z^{\prime}$ are cylindrical source coordinates
and $r$ and $z$ are the only observation point coordinates of relevance.
To evaluate Eq. (2) it is convenient to make use of the cylindrical
Green's function Bessel function expansion
$${1 \over \vert {\bf r} -{\bf r^{\prime}} \vert }=\sum_{m=-\infty}^{\infty}
\int_0^\infty dk J_m(kr)J_m(kr^{\prime})
e^{im(\phi-\phi^{\prime})-k \vert  z -  z^{\prime} \vert } \eqno(3)$$
Inserting Eq. (3) into Eq. (2) leads directly to
$$V_{\beta}(r,z)=-2\pi\beta\int_{0}^{\infty} dk\int_{0}^{\infty}dR
\int_{-\infty}^{\infty}dz^{\prime}
R \Sigma(R,z^{\prime})J_0(kr)J_0(kR) e^{-k\vert z -  z^{\prime}\vert}\eqno(4)$$
Finally, taking the disk to be
infinitely thin (viz. $\Sigma(R,z^{\prime})=\Sigma(R)\delta(z^{\prime})$)
then yields for points with $z=0$ the potential
$$V_{\beta}(r)=-2\pi\beta\int_{0}^{\infty} dk\int_{0}^{\infty}dR
R \Sigma(R)J_0(kr)J_0(kR) \eqno(5)$$
which we immediately recognize as Toomre's original result for an
infinitely thin disk. In passing we note that Eq. (4) also holds for
off-axis points with non-zero $z$, and actually provides
a generalization of Toomre's work to disks whose
thickness may not in fact be negligible, with the form of Eq. (4) being
particularly convenient if the fall-off of the matter distribution in the
$z$ direction is itself exponential.

For our purposes here, the expansion of Eq. (3) can immediately be
applied to the linear potential case too, and this leads directly
to the potential
$$V_{\gamma}(r,z)=$$
$$\pi\gamma\int dkdRdz^{\prime}R
\Sigma(R,z^{\prime})[(r^2+R^2+(z-z^{\prime})^2)J_0(kr)J_0(kR)
-2rR J_1(kr)J_1(kR)]
e^{-k\vert z -  z^{\prime}\vert} \eqno(6)$$
Equation (6) then  reduces at $z=0$ for infinitely thin disks to the
compact expression
$$V_{\gamma}(r)=\pi\gamma\int_{0}^{\infty} dk\int_{0}^{\infty}dRR
\Sigma(R)[ (r^2+R^2)J_0(kr)J_0(kR) -2rR J_1(kr)J_1(kR)]
\eqno (7)$$
If the $k$ integrations are performed first in Eqs. (5) and (7) they
lead to highly singular hypergeometric functions whose subsequent $R$
integrations contain infinities which, even when they are in fact
integrable, nonetheless require
great care when carried out numerically. However, as was noted by
Freeman (1970), in certain cases the $R$ integration may be carried out first,
a
can even yield a completely singularity free procedure. Indeed, for a thin
expon
disk with $\Sigma(R)=\Sigma_0exp(-\alpha R)$
where $R_0=1/\alpha$ is the scale height  and $N=2\pi\Sigma_0 R_0^2$
is the total number of particles in the disk, first the $R$ and then the $k$
integrations can be performed completely and analytically by use of standard
Bes
function integral formulas to then lead to Freeman's original result, viz.

$$V_{\beta}(r)= -\pi\beta\Sigma_0 r[I_0(\alpha r/2)K_1(\alpha r/2)-
I_1(\alpha r/2)K_0(\alpha r/2)]\eqno(8)$$
for the Newtonian contribution. A completely analogous treatment can be made
for
linear contribution, and following
some straightforward, albeit lengthy, algebra, we obtain for it the compact
expr

$$V_{\gamma}(r)= \pi\gamma\Sigma_0 \{ (r/\alpha^2)[I_0(\alpha r/2)K_1(\alpha
r/2
I_1(\alpha r/2)K_0(\alpha r/2)]$$
$$+ (r^2/2\alpha)[I_0(\alpha r/2)K_0(\alpha r/2)+
I_1(\alpha r/2)K_1(\alpha r/2)] \}
\eqno(9)$$

To obtain test particle rotational velocities  we need only differentiate
Eqs. (8) and (9) with respect to $r$. This is readily achieved via
repeated use of the modified Bessel function recurrence relations, and yields
$$rV^{\prime}(r)=
\pi\beta\Sigma_0\alpha r^2[I_0(\alpha r/2)K_0(\alpha r/2)-
I_1(\alpha r/2)K_1(\alpha r/2)]$$
$$+(\pi\gamma\Sigma_0 r^2/\alpha)I_1(\alpha r/2)K_1(\alpha r/2)
\eqno(10)$$
Using the asymptotic properties of the modified Bessel functions we find
that at distances much larger than the scale height $R_0$ Eq. (10) yields
$$rV^{\prime}(r) \rightarrow N\beta /r+
N\gamma r/ 2 -3N\gamma R_0^2/ 4 r \eqno(11)$$
as would be expected. The coefficient
$N\beta$ is usually identified as $MG/c^2$ with $M$ being taken to be the
mass of the disk. For normalization purposes it is convenient to use this
coefficient to define the velocity $v_0=(N\beta/R_0)^{1/2}$, the velocity
that a test particle would have if orbiting a Newtonian point galaxy with the
same total mass at a distance of one scale height. In terms of the convenient
parameter $\eta=\gamma R_0^2/\beta$, Eq. (10) then yields for the
rotational velocity $v(r)$ of a circular orbit in the plane of the exponential
disk the compact, exact expression
$$v^2(r)/v_0^2=(r^2\alpha^2/2)[I_0(\alpha r/2)K_0(\alpha r/2)+
(\eta-1)I_1(\alpha r/2)K_1(\alpha r/2)] \eqno(12)$$
which is our main result. All departures from the standard Freeman result
are thus embodied in the parameter $\eta$ in the simple manner indicated,
and we turn now to explore its consequences.

\noindent
{\bf (3) Exponential Disks and Flat Rotation Curves}

As a first attempt at fitting data we have chosen to study the rotation curve
of
the galaxy NGC3198, since for it the data go out to the largest known number of
surface brightness scale heights; and, with the data being so flat, this
galaxy is generally regarded as being prototypical.
To model the galaxy, we have followed van Albada et. al. (1985), and
represented the surface brightness by a single exponential with a $1^{\prime}$
($=2.72~kpc$) scale height. (This choice approximates Wevers et. al. (1986)
$U^{
$J$, and $F$ filter data with eyeball slopes of $R_0=63 ^{\prime \prime},~
58 ^{\prime \prime},~54 ^{\prime \prime}$ respectively at a $5\%$ uncertainty
level (the $F$ filter data have also been confirmed by Kent 1987),
while ignoring a  spike in the very small angle  region data, and also a
possible truncation at the edge of the visible region). Moreover, the model
igno
any modifications to the luminosity profile due to extinction or galactic dust
i
reprocessing. Following Begeman (1989)
we have assigned a $z-$thickness to the disk according to the general analysis
of van der Kruit and Searle (1981), so that the surface brightness function
$\Sigma(R, z^{\prime})$ takes the separable form $\Sigma(R)
sech^2(z^{\prime}/z_
/2z_0$ with $z_0=R_0/5$. Recognizing a $15\%$ or so contribution to the
visible mass density from the $HI$ gas itself, we have also included the gas as
a matter source, and have found that, for model purposes, Wevers et. al. (1986)
$HI$ surface density data can be well represented by a sum of three
exponentials
viz. $\sigma_{HI}(r)=(37.0exp(-0.45r)+34.6exp(-1.14r)-68.2exp(-0.83r)
)~M_{\odot}/pc^2$ with a total $HI$ mass (to infinity) of
$5.2\times10^{9}~M_{\o
of which $4.9(\pm 0.2)\times10^{9}~M_{\odot}$ is observed in the explored
$12^{\
region. Finally, again following Begeman, and also van Albada and Sancisi
(1986)
we have multiplied the $HI$ gas profile by a
factor of 1.4 to allow for the presence of helium. With the model thus defined,
we have used the method of Sec. (2) which has only the two free parameters
$\eta
$v_0$ of the stars to generate the fit of Fig. (1) to Begeman's (1989) rotation
data. With $v_0$
essentially being constrained by the overall normalization of the stellar
contri
our best fit is found to have a $\chi^2$ of 37.7 for the 28 data points, with
$\
taking the value 0.044 and $v_0$ the value $244.3~km/s$. Thus we obtain a value
$3.8\times10^{10}~M_{\odot}$ for the mass of the stars which is quite
reasonable for a galaxy with quoted luminosity
$L_B=(9.0\pm0.9)\times10^9L_{B\od
$L_V=(7.3\pm0.7)\times10^9L_{V\odot}$, with the obtained galactic mass being
a typical so called maximum disk mass
in which the Newtonian term gets to be as large as it possibly can be.
Additionally, from the value found for $\eta$ we deduce that the stellar
contribution to the linear term in Eq. (11) is given by
$1/\gamma_{galaxy}=1/N\g
{star}=2.9\times10^{29}~cm$. The linear term is thus competitive with the
Newton
in a galaxy when $1/\gamma_{galaxy}$
is of order the Hubble radius, an intriguing fact which had naively been
anticip
Mannheim and Kazanas (1989) in their original study. (In passing we note that
with such a small value
for $\gamma$ the $\beta\gamma$ product terms in Eq. (1) are then rendered
completely insignificant, a fact we had indicated earlier). Given this value of
$\gamma_{galaxy}$, a typical
value for $\gamma_{star}$ would then be of order $10^{-40}~cm^{-1}$, making the
linear
potential indeed negligible on solar system distance scales as initially
require
with the linear potential only first becoming competitive with the Newtonian
one on galactic distance scales.

It is important to stress that
the $z-$thickness structure of the disk is only significant at small radii
where
serves to ensure that the inner part of the rotation curve is well
fitted by the Newtonian contribution, to thus make it
possible to explore fully the effect of the linear term on the outer points.
It is interesting to note that the contribution of the linear potential piece
sh
in  Fig. (1) is
remarkably reminiscent in shape to that of a typical dark matter contribution
to
galactic data fitting (see e.g. Kent (1987) for an extensive study).
As we see, the obtained rotation curve has the characteristic
Newtonian driven rise in the region up to about 3 scale heights, and from
then onwards it is remarkably flat out to 11 scale heights. While we are always
able to adjust the parameter $\eta$ so that the Newtonian and linear terms will
just balance each other in some region, what may not immediately have been
anticipated prior to an explicit calculation is that the balancing region could
as large as the 8 or so scale heights we find, an enormous distance interval.
(T
hand side of Eq. (12) is simply a very slowly varying function in this region).
Over the range from 3 to 11 scale heights
the rotational velocity  $v(r)$ is found (in $km/s$) to take the values (156.3,
149.5, 147.3, 146.6, 147.3, 149.0, 151.3, 154.1 ) in unit step increases. Thus
i
has a spread of $\pm 3.2 \%$ about a central value of 151.5 in this region.
Additionally, we find that even at 15 scale heights
the velocity has still only increased to 167.7, an 11 per cent increase over
the
151.5 value. While our presented fit would appear to be acceptable and
at least have the right general structure, we should point out that at the
large
distances there are some not fully understood discrepancies (of order up to
$7~km/s$)  between Begeman's data and Bosma's earlier 1978, 1981 data, and
also we should note that Begeman's last 2 data points (the farthest) actually
us
values which are extrapolated from closer in ones. Also, there is even some
indi
in the data of a warp at the largest observed distances which we have not
attemp
model. Further, the inner $2^{\prime}$ region has also been explored by Hunter
e
al. (1986), with Begeman quoting a maximum difference between the two inner
rota
curve data sets of $3~km/s$. Consequently, it is not immediately clear what
leve
precision fitting of data is currently warranted (for instance, a simple
$1~km/s
addition to each of Begeman's quoted errors and no refinement to our surface
brightness model brings the best $\chi^2$ down to 19.7); and all in all we do
fe
our fitting does give a first measure of
encouragement for the possible presence of a linear gravitational potential
term

In order to assess the possible significance of our analysis, it is
important to clarify the meaning of the term 'flat rotation curve'. In the
literature it is generally thought that rotation curves will be flat
asymptotica
(though of course the more significant issue is the fact that they deviate from
luminous Newtonian prediction at all, rather than in what particular way); and
o
since our model predicts that velocities will
eventually grow as $r^{1/2}$, the initial expectation is that the model
is immediately ruled out. However, the fits that have so far been made are
not in fact asymptotic ones. Firstly, the $HII$ optical studies pioneered by
Rubin et. al. (1978), even while they were indeed yielding flat rotation
curves,
were restricted to the somewhat closer in optical disk region.
And eventually, after a concentrated
effort to carefully measure the surface brightness of the such galaxies,
it was found (see e.g. Kalnajs (1983), Kent (1986)) that
the $HII$ curves could be described, albeit coincidentally, by a
standard luminous Newtonian prediction (except perhaps for
the very last few observed points); even in fact for galaxies such as UGC2885
for which the data go out to as much as $80~kpc$, simply because the optical
dat
do not (because of their very nature) go out to a
large enough number of scale heights. As to $HI$ gas studies which do indeed
probe the farther out reaches of a galaxy, here the standard dark matter fits
(van Albada et. al. (1985), Kent (1987)) achieve flatness not through the
flatness of a pure (logarithmic) dark matter component, but rather through an
in
(the so called conspiracy recognized by van Albada and Sancisi (1986) and
explored recently by Casertano and van Gorkom (1991)) between a
falling Newtonian piece and a rising dark matter piece, in a manner
analogous to the interplay found in the present work in Fig. (1),
to again not be indicative of true asymptotic dynamics. Thus, for the moment,
even though both the available $HI$ and $HII$ type data are flat in their
respective domains, each data set is flat for its own coincidental reason, and
it would appear to us that true galactic asymptotia has yet to be
explored; with the observed flatness of the galactic rotation curves
(just like the apparent flatness of total proton proton scattering cross
section
many energy decades before an eventual rise) perhaps only being an
intermediate rather than an asymptotic phenomenon.

In the spirit of trying to work up from data to establish the gravitational
potential, we also recall that one of the original motivations for dark matter
was the need to stabilize Newtonian galaxies (Ostriker and Peebles
(1973)). It is therefore quite interesting to note that numerical
simulations based on our Newtonian plus linear potential and luminous matter
alo
(Christodoulou
(1991)) have also been found to lead to galactic stability. A systematic,
overall multi-galactic study of stability and rotational velocities which
uses the same
law of force and matter distributions for both of these galactic issues
simultaneously would thus appear to be worthwhile.
Beyond galactic distance scale dynamics, perhaps
the sharpest difference between the second and the fourth order theories may
eme
the slightly bigger scale associated with
gravitational lensing where differences between linear and logarithmic
potential
even become pronounced, thus making a study of
the (so far unknown) conformal gravity lensing
predictions in the non-asymptotically flat geometry of Eq. (1) quite urgent.

As a final remark on the Newtonian limit, we recall that the exterior
Newtonian potential goes hand in hand with the interior second order
Poisson equation in the standard picture, this being another cornerstone
of the standard second order non-relativistic gravitational theory.
Now in their study of the conformal theory, Mannheim and Kazanas (1992) have
fou
the interior problem reduces exactly and without approximation to a fourth
order
equation, with the metric of Eq. (1) with its explicit Newtonian term being
none
than its solution.
Thus while the second order Poisson
equation implies the Newtonian potential, it is not true that
Newton's Law of Gravity implies the standard second order Poisson
equation. The second order Poisson equation (and accordingly its second order
Einstein covariant generalization) are thus not in fact necessary
but only sufficient ingredients
to yield Newton's Law, thereby opening up the possibility
of other candidate theories of gravity such as the one considered by
Mannheim and Kazanas; with differentiation between them only being
decidable in the final analysis by observation. We believe that we have
thus established the candidacy (at least)
of fourth order gravity by working up from the
non-relativistic limit; and since the conformal theory has already been
shown to possess no flatness problem (Mannheim (1992)) and thus not require
any cosmological dark matter, we see that both cosmologically and
galactically it might turn out to be the case that luminous matter is the
major constituent of the Universe after all.

The author would like to thank Demos Kazanas, Dimitri Christodoulou and Jason
Ta
for stimulating discussions. This work has been supported in part
by the Department of Energy under grant No. DE-FG02-92ER40716.00.
\vfill\eject
\noindent
{\bf References}
\medskip
\noindent Begeman, K. G. 1989, A. A., 223, 47.
\smallskip
\noindent Bosma, A. 1978, Ph. D. Thesis, Gronigen University, unpublished.
\smallskip
\noindent Bosma, A. 1981, A. J., 86, 1791.
\smallskip
\noindent Casertano, S., and van Gorkom, J. H. 1991, A. J., 101, 1231.
\smallskip
\noindent Christodoulou, D. M. 1991, Ap. J., 372, 471.
\smallskip
\noindent Freeman, K. C. 1970, Ap. J. 160, 811.
\smallskip
\noindent Hunter, D. A., Rubin, V. C., and Gallagher III, J. S. 1986, A. J. 91,
\smallskip
\noindent Kalnajs, A. J. 1983, in Internal Kinematics and Dynamics of Disk
Galax
IAU Symposium No. 100, ed. E. Athanassoula (Reidel, Dordrecht), p.87.
\smallskip
\noindent Kazanas, D., and Mannheim, P. D. 1991, Ap. J. Suppl.
Ser., 76, 431.
\smallskip
\noindent Kent, S. M. 1986, A. J. 91, 1301.
\smallskip
\noindent Kent, S. M. 1987, A. J. 93, 816.
\smallskip
\noindent Mannheim, P. D. 1990, Gen. Rel. Grav., 22, 289.
\smallskip
\noindent Mannheim, P. D. 1992, Ap. J., 391, 429.
\smallskip
\noindent Mannheim, P. D., and Kazanas, D. 1989, Ap. J., 342, 635.
\smallskip
\noindent Mannheim, P. D., and Kazanas, D. 1991, Phys. Rev., D44, 417.
\smallskip
\noindent Mannheim, P. D., and Kazanas, D. 1992, Newtonian limit of
conformal gravity and the lack of necessity of the second order Poisson
equation, NASA Goddard preprint, unpublished.
\smallskip
\noindent Milgrom, M. 1983, Ap. J. 270, 365; 371, 384.
\smallskip
\noindent Ostriker, J. P., and Peebles, P. J. E. 1973, Ap. J. 186, 467.
\smallskip
\noindent Rubin, V. C., Ford W. K., and Thonnard, N. 1978, Ap. J. (Letters),
225, L107.
\smallskip
\noindent Sanders, R. H. 1990, A. A. Rev., 2, 1.
\smallskip
\noindent Toomre, A. 1963, Ap. J., 138, 385.
\smallskip
\noindent van Albada, T. S., Bahcall, J. N., Begeman, K. G., and Sancisi,
R., 1985, Ap. J., 295, 305.
\smallskip
\noindent van Albada, T. S., and Sancisi, R.,
1986, Phil. Trans. R. Soc., A320, 447.
 \smallskip
\noindent  van der Kruit, P. C., and Searle L., 1981,
A. A., 95, 105.
 \smallskip
\noindent Wevers, B. M. H. R., van der Kruit, P. C., and Allen, R. J. 1986,
A. A. Suppl. Ser., 66, 505.
\bigskip
\noindent
{\bf Figure Caption}

\medskip
\noindent Figure (1). Comparison of the calculated rotational velocity curve
for
NGC3198 with Begeman's data shown with his quoted error bars. The full
curve shows the overall theoretical velocity prediction in $km/s$,
while the two indicated dotted curves show the rotation curves that
separate Newtonian and linear potentials would produce.
\end